\documentclass[12pt,letter]{article}

\usepackage{graphicx, epsfig, color,cite}
\usepackage{amsmath}
\usepackage{amssymb}
\usepackage{subfigure}

\def\eslt{E_T^{\rm miss}}
\def\delew{\Delta_{\rm EW}}

\def\to{\rightarrow}

\def\bi{\begin{itemize}}
\def\ei{\end{itemize}}

\def\sps1ap{SPS1a$^\prime$}
\def\c1p{C1$^\prime$}

\def\tb{\tilde b}

\def\tst{\tilde t}
\def\ttau{\tilde \tau}

\def\tg{\tilde g}

\def\tq{\tilde q}
\def\tw{\widetilde W}
\def\tz{\widetilde Z}
\def\alt{\lesssim}
\def\agt{\gtrsim}
\def\be{\begin{equation}}  
\def\ee{\end{equation}}  
\def\bea{\begin{eqnarray}}  
\def\eea{\end{eqnarray}}  
\def\beas{\begin{eqnarray*}}  
\def\eeas{\end{eqnarray*}}  
\newcommand\prd[3]{{\it Phys.\ Rev.\ }{\bf D #1} (#2) #3}

\newcommand\plb[3]{{\it Phys.\ Lett.\ }{\bf B #1} (#2) #3}

\newcommand\npb[3]{{\it Nucl.\ Phys.\ }{\bf B #1} (#2) #3}

\begin{document}
\begin{titlepage}
\begin{flushright}
OU-HEP-161104 \\
UT-16-32 \\
\end{flushright}

\vspace{0.5cm}
\begin{center}
{\LARGE \bf A top-squark hunter's guide
}\\ 
\vspace{1.2cm} \renewcommand{\thefootnote}{\fnsymbol{footnote}}
{\large 
Howard Baer$^{1}$\footnote[1]{Email: baer@ou.edu },
Vernon Barger$^{2}$\footnote[2]{Email: barger@pheno.wisc.edu },
Natsumi Nagata$^{3}$\footnote[3]{Email: natsumi@hep-th.phys.s.u-tokyo.ac.jp}
and Michael Savoy$^1$\footnote[4]{Email: savoy@nhn.ou.edu}
}\\ 
\vspace{1.2cm} \renewcommand{\thefootnote}{\arabic{footnote}}
{\it 
$^1$Department of Physics and Astronomy, \\
University of Oklahoma, Norman, OK 73019, USA \\[3pt]
}
{\it 
$^2$Department of Physics and Astronomy, \\
University of Wisconsin, Madison, WI 53706, USA \\[3pt]
}
{\it 
$^3$Department of Physics, University of Tokyo, Bunkyo-ku, Tokyo 113-0033, 
Japan \\[3pt]
}

\end{center}

\vspace{0.5cm}
\begin{abstract}
\noindent 
In supersymmetric models with radiatively-driven naturalness and light
higgsinos, the top squarks may lie in the 0.5--3~TeV range and thus
only a fraction of natural parameter space is accessible to LHC searches. 
We outline the range of top squark and lightest SUSY particle 
masses preferred by electroweak naturalness in the standard
parameter space plane.
We note that the branching fraction for $b\to s\gamma$ decay favors top squarks
much heavier than 500~GeV. 
Such a range of top-squark mass values is in contrast to previous expectations
where $m({\rm stop})<500$~GeV had been considered natural.
In radiative natural SUSY, top squarks decay roughly equally via $\tst_1\to b\tw_1$ and 
$t\tz_{1,2}$ where $\tw_1$ and $\tz_{1,2}$ are higgsino-like electroweak-inos. 
Thus, top squark pair production should yield all of 
$t\bar{t}+\eslt$, $t\bar{b}+\eslt$, $b\bar{t}+\eslt$ and
 $b\bar{b}+\eslt$ signatures at comparable rates. 
We propose that future LHC top squark searches take place within a 
semi-simplified model which corresponds more closely to expectations 
from theory.

\vspace*{0.8cm}

\end{abstract}

\end{titlepage}

\section{Introduction}
\label{sec:intro}

The supersymmetrized (SUSY) Standard Model (SM), \textit{e.g.}, the minimal
supersymmetric Standard Model (MSSM), has for a long time intrigued
particle theorists in that it is free of the scalar field quadratic
divergences that plague non-supersymmetric theories \cite{witten}.
In addition, the MSSM has made three predictions which have since been
verified by experiment: 1. the value of $\sin^2\theta_W\simeq 0.232$
which arises from unified gauge couplings at $m_{\rm GUT} \simeq 2\times
10^{16}$~GeV that evolve via renormalization group (RG) evolution down to the
weak scale within the context of the MSSM \cite{gauge}, 
2. the large top quark mass $m_t\simeq 173$~GeV \cite{rewsb} is exactly
what is needed to initiate a radiative breakdown of electroweak symmetry
in the MSSM \cite{rewsb}, and 3. the measured value of the Higgs boson
mass $m_h\simeq 125$~GeV \cite{atlas_h,cms_h, Aad:2015zhl} which falls
squarely within the narrow window required by the MSSM \cite{mhiggs}.

In contrast, so far no evidence for direct production of superpartners
has emerged at LHC, leading to mass limits $m_{\tg}\agt 1900$~GeV
\cite{atlas_gl, CMS:2016xva} and $m_{\tst_1}\agt 850$~GeV
\cite{atlas_stop, atlas_stop_run2, CMS:2016vew} in the context of various simplified
models. The latter lower bound has been particularly disconcerting since
it is in direct conflict with an oft-repeated mantra that one or more
light third generation squarks ($m_{\tst_1}\alt 500$~GeV) are required
for a {\it natural} SUSY solution to the Little Hierarchy (LH)
problem. Here, the LH is characterized by the growing gap between the
weak scale, as represented by $m_{W,Z,h}\sim 100$~GeV, and the
superparticle mass scale $m_{\rm SUSY}$ which apparently lies within the
multi-TeV range.

The light top squark narrative has lead to an ``all hands on deck'' call for 
exploring every conceivable gap of allowed masses and decay modes in the
simplified model $m_{\tst_1}$ vs. $m({\rm LSP})$ (the LSP, lightest SUSY
particle) plane. The impression has been made that by covering every
possibility for existence of light top squarks, then one may be ruling
out weak scale SUSY or else showing that whatever form SUSY takes, it is
not as ``we'' understood it \cite{craig}. The top squark mass bound is
also being invoked to justify costly decisions regarding future
experimental facilities: if weak scale SUSY as we know it is ruled out,
and the SM remains valid well into the multi-TeV range, then perhaps a
100~TeV hadron collider is the way to go as all bets from theory would
be off. Alternatively, if SUSY remains just beyond the energy horizon,
then perhaps ILC and an energy upgrade LHC (HE-LHC) operating with
$\sqrt{s}\sim 28$--33~TeV are the right machines to build. Given the
stakes involved, it is becoming critical to ensure the validity of our
reasoning regarding the notions of electroweak naturalness and
fine-tuning.

To address this issue, in Sec.~\ref{sec:nat} we briefly review several
estimates of electroweak naturalness in the SM and in SUSY. We believe
that several common measures are technically mis-applied in the SUSY
case. When corrected to allow for the fact that the
soft parameters should be correlated, 
they reduce to the model independent measure
$\Delta_{\rm EW}$, where EW denotes electroweak\cite{ltr, rns}. 
The latter measure also leads to bounds on top squarks and gluinos,
but instead allows for $m_{\tst_1}\alt 3$~TeV and $m_{\tg}\alt 4$~TeV at
little cost to naturalness since these masses enter into the value of
$m_Z$ as finite one- and two-loop corrections respectively. In
Sec.~\ref{sec:BM}, we present a top squark benchmark model from the
two-extra-parameter non-universal Higgs model \cite{nuhm2} (NUHM2) which
allows for highly natural SUSY spectra with $m_h\simeq 125$~GeV. This
leads to a grand overview plot of expectations for populating the
$m_{\tst_1}$ vs. $m_{\tz_1}$ plane in Sec.~\ref{sec:plane}. This plot
presents a guide for top squark hunters at the LHC as to where in the
plane their quarry of natural SUSY solutions lies for low values of
$\Delta_{\rm EW}$. Here, we find $m_{\tst_1}\alt 1.2$--1.8~TeV for
$\Delta_{\rm EW}<15$ while $m_{\tst_1}\alt 3$~TeV for $\Delta_{\rm
EW}<30$. Hardly any solutions lie in the highly scrutinized {\it
compressed} region where $m_{\tst_1}\sim m_{\tz_1}$. In
Sec.~\ref{sec:bsg}, we evaluate expectations for the flavor-changing
decay ${\rm BF}(b\to s\gamma )$ versus $m_{\tst_1}$ and find for
$m_{\tst_1}<500$~GeV that one always expects large deviations from the
measured value whereas for $m_{\tst_1}>1.5$~TeV, then the SUSY loops
decouple and one gains accord with experiment: in this sense, it comes
as no great surprise that LHC top squark hunters have yet to sight their
trophy. In Sec.~\ref{sec:prod}, we outline top squark production and
decay rates for natural SUSY and in Sec.~\ref{sec:prospects} we outline
a more realistic proposal for future top squark searches in a
semi-simplified model which corresponds more closely with predictions
from theory. A summary and conclusions are given in
Sec.~\ref{sec:conclude}.\footnote{Some early work on top squark
phenomenology is given in Ref's~\cite{Bigi:1985aq, Baer:1985hd,
Hikasa:1987db, Drees:1990te, Baer:1991cb, Olive:1994qq}. Some recent
examinations include Ref's~\cite{Meade:2006dw, Graesser:2008qi, Drees:2012dd, Cao:2012rz, Bai:2012gs, Han:2012fw, Han:2013kga,
Bai:2013xla, Demir:2014jqa, Czakon:2014fka, Kobakhidze:2015scd,
Abe:2015xva, Belanger:2015vwa, An:2015uwa, Hikasa:2015lma,
Rolbiecki:2015lsa, Belyaev:2015gna, Macaluso:2015wja, Kawamura:2016drh,
Goncalves:2016tft,
Han:2016hgr, Han:2016xet, Cheng:2016mcw, Buckley:2016kvr,
Pierce:2016nwg, Bai:2016zou, Duan:2016vpp, Cici:2016oqr}.}

\section{Brief review of naturalness}
\label{sec:nat}

\subsection{Fine-tuning rule}

For any observable ${\cal O}$, if the contributions to ${\cal O}$ are given by
\be
{\cal O}= a +b +f(b) +c ~,
\ee
then we would claim the value of ${\cal O}$ is {\it natural} if each
contribution on the right-hand-side is comparable to or less than ${\cal
O}$. If this were not the case, if say one contribution $c$ were far
larger than ${\cal O}$, then some other contribution would have to be
fine-tuned to large opposite-sign values such as to maintain the
measured value of ${\cal O}$. Thus, the naturalness measure  
\be
\Delta =|\text{largest contribution to RHS}|/|{\cal O}|
\ee 
would be vindicated (here, RHS stands for right-hand-side). 
In the case of the quantity $f(b)$, if as a
consequence of $b$ getting large, then $f(b)$ becomes large negative,
these two quantities are {\it dependent} and should be {\it combined}
before evaluating naturalness. This is embodied by the fine-tuning rule
articulated in Ref. \cite{dew}: {\it in evaluating fine-tuning, it is
not permissible to claim fine-tuning of {\rm dependent} quantities one
against another}.

\subsection{Higgs mass fine-tuning in the SM}

For illustration, in the case of the SM with a scalar potential given by 
\be
V=-\mu^2_{\rm SM}|\phi^\dagger \phi |+\lambda |\phi^\dagger\phi|^2,
\ee
 the physical Higgs boson mass is given by
\be
m_{H_{\rm SM}}^2 \simeq  2\mu^2_{\rm SM}+\delta m_{H_{\rm SM}}^2,
\ee
where the largest contribution to $\delta m_{H_{\rm SM}}^2$ comes from
the famous quadratic divergences: 
\be
\delta m_{H_{\rm SM}}^2\simeq
\frac{3}{4\pi^2}\left(-\lambda_t^2+\frac{g^2}{4}
+\frac{g^2}{8\cos^2\theta_W}+\lambda\right) 
\Lambda^2,  
\ee
where $\lambda_t$ is the SM top-quark Yukawa coupling, $g$ is the
$SU(2)_L$ gauge coupling and $\Lambda$ represents the energy scale
cut-off on the quadratically divergent one-loop mass corrections. Since
$2\mu_{\rm SM}^2$ is {\it independent} of $\delta m_{H_{\rm SM}}^2$,
then $\mu_{\rm SM}^2$
can be freely dialed, or fine-tuned, to maintain the measured value of
$m_{H_{\rm SM}}=125.1$~GeV \cite{Aad:2015zhl}. A valid measure of
fine-tuning here would be $\Delta_{\rm SM}=|\delta m_{H_{\rm
SM}}^2|/m_{H_{\rm SM}}^2$. Requiring $\Delta_{\rm SM}< 30$ implies  an
upper bound on the SM effective theory energy cutoff of $\Lambda\alt
5.8$~TeV. 

\subsection{Higgs mass fine-tuning in the MSSM}

The situation in the MSSM is quite different \cite{comp3}. 
In this case, the well-known quadratic divergences all cancel but there
remains a variety of intertwined logarithmic divergent contributions to
$m_h^2$. In the MSSM, we have 
\be
m_h^2\simeq -2\left\{ \mu^2(\text{weak}) +m_{H_u}^2(\text{weak})\right\}
\sim -2\left\{
\mu^2(\Lambda ) +m_{H_u}^2(\Lambda ) +\delta m_{H_u}^2(\Lambda )
\right\}~,
\ee
where now $\mu$ is the superpotential higgsino mass term and $m_{H_u}^2$ is 
the up-Higgs soft SUSY breaking squared mass. 
The quantity $\delta m_{H_u}^2$ is properly evaluated by integrating the renormalization group
equation: 
\be 
\frac{dm_{H_u}^2}{dt}= \frac{2}{16\pi^2}\left(-\frac{3}{5}g_1^2M_1^2-3g_2^2M_2^2+\frac{3}{10}g_1^2S+3f_t^2X_t\right),
\ee 
where $t=\log (Q)$ with $Q$ the renormalization scale, $M_i$ ($i=1$--3)
are the various gaugino masses, $g_i$ are the corresponding
gauge coupling constants, $f_t$ is the top Yukawa coupling, 
\be
S=m_{H_u}^2-m_{H_d}^2+{\rm Tr}\left[ {\bf m}_Q^2-{\bf m}_L^2-2{\bf
m}_U^2+{\bf m}_D^2+{\bf m}_E^2 \right] ~,
\ee
 and 
\be
X_t=m_{Q_3}^2+m_{U_3}^2+m_{H_u}^2+A_t^2 ~,
\label{eq:Xt}
\ee
where $m_{H_d}^2$, ${\bf m}_Q^2$, ${\bf m}_L^2$, ${\bf m}_U^2$, ${\bf
m}_D^2$, ${\bf m}_E^2$ are the soft masses for the down-type Higgs,
left-handed squarks, left-handed sleptons, right-handed up-type squarks,
right-handed down-type squarks, and right-handed charged sleptons,
respectively, and $A_t$ is the $A$-term for the top Yukawa coupling. 
To evaluate $\delta m_{H_u}^2$, it is common in the literature to set
the gauge couplings, the $S$ parameter and $m_{H_u}^2$ equal to zero so
that a simple one step integration can be performed leading to 
\be
\delta m_{H_u}^2(\Lambda)
\sim -\frac{3 f_t^2}{8\pi^2}
\left(m_{Q_3}^2+m_{U_3}^2+A_t^2\right) \log\left(\frac{\Lambda}
{m_{\rm SUSY}}\right) 
~.
\label{eq:delmHus}
\ee
The fine-tuning measure $\Delta_{\rm HS}=|\delta m_{H_u}^2|/m_h^2\alt 30$
requires at least one (and actually three) third generation squarks with
mass less than 650~GeV\cite{oldnsusy}.

The issue here is that, unlike the SM case, $\delta m_{H_u}^2(\Lambda )$
is {\it not independent} of the high scale value of $m_{H_u}^2(\Lambda
)$. In fact, the larger $m_{H_u}^2 (\Lambda )$ is, the larger is the
cancelling correction $\delta m_{H_u}^2$. This violates the fine-tuning
rule.

Instead, one ought to first {\it combine} dependent contributions, then
evaluate the independent contributions to the observed value of $m_h^2$
to check whether they exceed its value. Upon regrouping $m_h^2= -2
\left\{\mu^2+(m_{H_u}^2(\Lambda )+\delta m_{H_u}^2(\Lambda ))\right\}$,
where $m_{H_u}^2(\Lambda )+\delta m_{H_u}^2(\Lambda)=m_{H_u}^2({\rm
weak})$. Then it is seen that the criteria for naturalness is that the
{\it weak scale} values of $\mu^2$ and $m_{H_u}^2$ are each comparable
to $m_h^2$. This corrected measure allows for {\it radiatively-driven
naturalness}: large, unnatural values of $m_{H_u}^2$ at the high scale
$\Lambda$ may be driven to natural values at the weak scale via
radiative corrections\cite{ltr,rns}.

\subsection{BG fine-tuning: multiple or just one soft parameter?}

The measure $\Delta_{\rm BG}\equiv {\rm max}_i |\frac{\partial\log
m_Z^2}{\partial\log p_i} |$ was proposed by Ellis {\it et al.}
\cite{eenz} and investigated more thoroughly by Barbieri and Giudice
\cite{bg}. Here, the $p_i$ are fundamental parameters of the theory
labeled by index $i$. To begin, one may express $m_Z^2$ in terms of weak
scale SUSY parameters 
\be
m_Z^2 \simeq -2\mu^2 -2m_{H_u}^2 ~,
\label{eq:mZsapprox}
\ee
where the partial equality holds for moderate-to-large $\tan\beta$
values ($\tan \beta \equiv \langle H_u \rangle /
\langle H_d \rangle$ is the ratio of the Higgs VEVs) and where we assume for now the radiative corrections are
small. Next, one needs to know the explicit dependence of $m_{H_u}^2$
and $\mu^2$ on the fundamental parameters. Semi-analytic solutions to
the one-loop renormalization group equations for $m_{H_u}^2$ and $\mu^2$
can be found for instance in Ref's~\cite{munoz}. For the case of
$\tan\beta =10$, it is found that \cite{abe, martin, feng} 
\bea
m_Z^2& \simeq & -2.18\mu^2 + 3.84 M_3^2+0.32M_3M_2+0.047 M_1M_3-0.42 M_2^2 \nonumber \\
& & +0.011 M_2M_1-0.012M_1^2-0.65 M_3A_t-0.15 M_2A_t\nonumber \\
& &-0.025M_1 A_t+0.22A_t^2+0.004 M_3A_b\nonumber \\
& &-1.27 m_{H_u}^2 -0.053 m_{H_d}^2\nonumber \\
& &+0.73 m_{Q_3}^2+0.57 m_{U_3}^2+0.049 m_{D_3}^2-0.052 m_{L_3}^2+0.053 m_{E_3}^2\nonumber \\
& &+0.051 m_{Q_2}^2-0.11 m_{U_2}^2+0.051 m_{D_2}^2-0.052 m_{L_2}^2+0.053 m_{E_2}^2\nonumber \\
& &+0.051 m_{Q_1}^2-0.11 m_{U_1}^2+0.051 m_{D_1}^2-0.052 m_{L_1}^2+0.053 m_{E_1}^2 ~,
\label{eq:mZsparam}
\eea
where all terms on the right-hand-side are understood to be $GUT$ scale
parameters.

The conundrum is then: what constitutes fundamental parameters? If all
GUT scale parameters on the RHS of Eq.~\eqref{eq:mZsparam} are
fundamental, then for the doublet top squark soft term we would find
$\Delta_{\rm BG} \sim 0.73 m_{Q_3}^2/(m_Z^2/2)$ and so $\Delta_{\rm
BG}<30$ would imply $m_{Q_3}\alt 400$~GeV in accord with
Eq.~\eqref{eq:delmHus}.

If instead we assume scalar mass universality as in the CMSSM, then the
fourth and fifth lines of Eq.~\eqref{eq:mZsparam} combine to $0.027
m_0^2$ and instead $\Delta_{\rm BG}=0.027 m_0^2/(m_Z^2/2)<30$ would
require $m_0\alt 2$~TeV: multi-TeV scalars are natural as in focus-point
SUSY \cite{Feng:1999mn}.

In fact, in more fundamental supergravity theories with SUGRA breaking
in a hidden sector, then {\it all} soft terms are computable as
multiples of the more fundamental gravitino mass $m_{3/2}$
\cite{sw}. Then all soft terms on the RHS of Eq.~\eqref{eq:mZsparam} are
{\it dependent} and must be combined according to the fine-tuning
rule. In this case, Eq.~\eqref{eq:mZsparam} collapses to a simpler form
\cite{dew}: 
\be
m_Z^2\simeq -2.18\mu^2 +a\cdot m_{3/2}^2 ~,
\label{eq:mzsm32}
\ee
and instead low fine-tuning requires $\mu\sim m_Z$ and also
$\sqrt{|a\cdot m_{3/2}^2|}\sim m_Z$. Equating Eq.~\eqref{eq:mZsapprox}
with Eq.~\eqref{eq:mzsm32} shows that $a\cdot m_{3/2}^2\sim
-m_{H_u}^2({\rm weak})$ and so we are led to consistency with the
corrected implication of $\Delta_{\rm HS}$: the criteria for electroweak
naturalness is that the weak scale values of $|m_{H_u}|$ and $|\mu |$
are $\sim m_{W,Z,h}\sim 100$~GeV.

\subsection{The electroweak measure $\Delta_{\rm EW}$}

The corrected versions of $\Delta_{\rm HS}$ and $\Delta_{\rm BG}$ are
consistent with requiring low {\it electroweak} fine-tuning in $m_Z^2$.
Minimization of the scalar potential in the minimal supersymmetric
Standard Model (MSSM) leads to the well-known relation \cite{wss}
\bea
\frac{m_Z^2}{2}&=&\frac{m_{H_d}^2+\Sigma_d^d-(m_{H_u}^2+\Sigma_u^u)\tan^2\beta}{\tan^2\beta -1}-\mu^2\label{eq:mzs1} \\
&\simeq& -m_{H_u}^2-\Sigma_u^u-\mu^2,
\label{eq:mzs}
\eea
where $\Sigma_u^u$ and $\Sigma_d^d$ denote the 1-loop corrections
(expressions can be found in the Appendix of Ref.~\cite{rns}) to the
scalar potential, $m_{H_u}^2$ and $m_{H_d}^2$ are the Higgs soft masses
at the weak scale. 
The second line is obtained from moderate to large values of $\tan\beta
\agt 5$ (as required by the Higgs mass calculation\cite{mhiggs}).
SUSY models requiring large cancellations between the various terms on the
right-hand-side of Eq.~\eqref{eq:mzs} to reproduce the measured value of
$m_Z^2$ are regarded as unnatural, or fine-tuned. In contrast, SUSY
models which generate terms on the RHS of Eq.~\eqref{eq:mzs} which are
all less than or comparable to $m_{\rm weak}$ are regarded as
natural. Thus, the {\it electroweak} naturalness measure $\delew$ is
defined as \cite{ltr,rns} 
\be
\delew\equiv \text{max}|{\rm each\ additive\ term\ on\ RHS\ of\
Eq.}~\eqref{eq:mzs1}|/(m_Z^2/2). 
\ee
Including the various radiative corrections, over 40 terms
contribute. The measure $\Delta_{\rm EW}$ is programmed in the Isajet
spectrum generator Isasugra \cite{isajet}. Neglecting radiative corrections, and taking moderate-to-large $\tan\beta \gtrsim 5$, 
then $m_Z^2/2 \sim-m_{H_u}^2-\mu^2$ so the main
criterion for naturalness is that {\it at the weak scale} 
\bi
\item $m_{H_u}^2\sim -m_Z^2$ and 
\item $\mu^2\sim m_Z^2$ \cite{ccn}.
\ei
The value of $m_{H_d}^2$ (where $m_A\sim m_{H_d}(\text{weak})$ with
$m_A$ being the mass of the CP-odd Higgs boson) can lie in the TeV$^2$ range
since its contribution to the RHS of Eq.~\eqref{eq:mzs} is suppressed by
$1/\tan^2\beta$. The largest radiative corrections typically come from 
the top squark sector:
\be
\Sigma_u^u (\tst_{1,2})= \frac{3}{16\pi^2}F(m_{\tst_{1,2}}^2)\times
\left[ f_t^2-g_Z^2\mp \frac{f_t^2 A_t^2-8g_Z^2(\frac{1}{4}-\frac{2}{3}\sin^2\theta_W)\Delta_t}{m_{\tst_2}^2-m_{\tst_1}^2}
\right]~,
\label{eq:Siguu}
\ee
where $\theta_W$ is the weak mixing angle, 
$\Delta_t=(m_{\tst_L}^2-m_{\tst_R}^2)/2+M_Z^2\cos
2\beta(\frac{1}{4}-\frac{2}{3}\sin^2\theta_W)$, $g_Z^2=(g^2+g^{\prime 2})/8$,
and $F(m^2)=m^2\left(\log
(m^2/Q^2)-1\right)$, with $Q^2=m_{\tst_1}m_{\tst_2}$.
Requiring highly mixed TeV-scale top squarks minimizes
$\Sigma_u^u(\tst_{1,2})$ whilst lifting the Higgs mass $m_h$ to $\sim
125$~GeV \cite{rns}. 

Using $\Delta_{\rm EW}<30$ or better than 3\% fine-tuning\footnote{For
higher values of $\Delta_{\rm EW}$, high fine-tuning sets in and is
displayed visually in Fig.~2 of Ref.~\cite{upper}. } 
then instead of earlier upper bounds, it is found that
\bi
\item $m_{\tg}\alt 4$~TeV,
\item $m_{\tst_1}\alt 3$ TeV and
\item $m_{\tw_1,\tz_{1,2}}\alt 300$ GeV.
\ei
Thus, gluinos and squarks may easily lie beyond the current reach of LHC
at little cost to naturalness while only the higgsino-like lighter
charginos and neutralinos are required to lie near the weak scale. 
The lightest higgsino $\tz_1$ comprises a portion of the dark
matter and would escape detection at LHC. 
The remaining dark matter abundance might be comprised of {\it e.g.} axions\cite{Bae:2013bva}.
Owing to their compressed
spectrum with mass gaps $m_{\tw_1}-m_{\tz_1}\sim m_{\tz_2}-m_{\tz_1}\sim
10$--20~GeV, the heavier higgsinos are difficult to see at LHC owing to
the rather small visible energy released from their three body decays
$\tw_1\to f\bar{f}'\tz_1$ and $\tz_2\to f\bar{f}\tz_1$ (where the $f$
stands for SM fermions).

\section{Illustration from a SUSY benchmark model}
\label{sec:BM}

In this section, we illustrate some aspects of top squark and Higgs boson masses
and mixings for a sample SUSY benchmark model from the two-extra-parameter
non-universal Higgs model (NUHM2\cite{nuhm2}) with parameter 
space given by
\be
m_0,\ m_{1/2},\ A_0,\ \tan\beta ,\ \mu,\ m_A,
\ee
where the non-universal GUT scale parameters $m_{H_u}^2$ and $m_{H_d}^2$ have
been exchanged for the more convenient weak scale values of $\mu$ and $m_A$.
Here, we will adopt parameter choices $m_0=5$ TeV, $m_{1/2}=900$ GeV, 
$\tan\beta =10$, $\mu =125$ GeV and $m_A=1$ TeV. 

%
\begin{figure}[tbp]
\begin{center}
 \includegraphics[clip, width = 0.49 \textwidth]{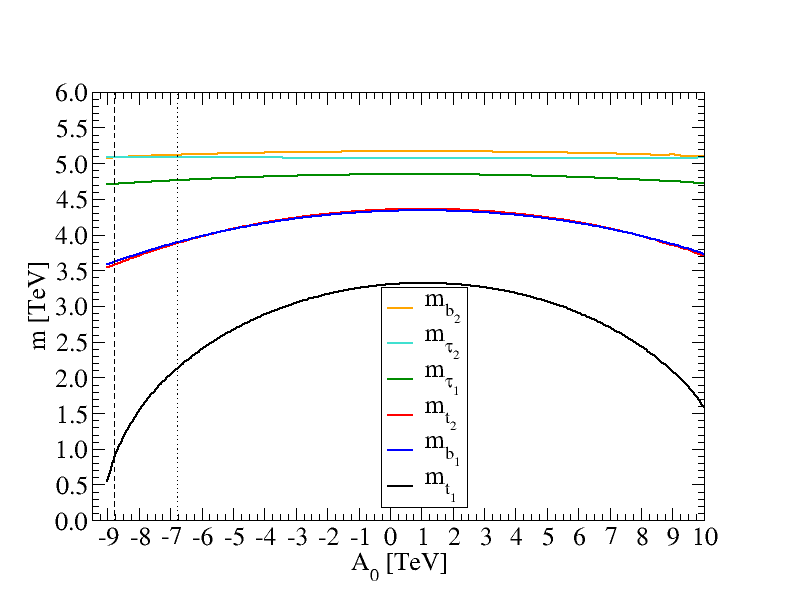}
 \includegraphics[clip, width = 0.49 \textwidth]{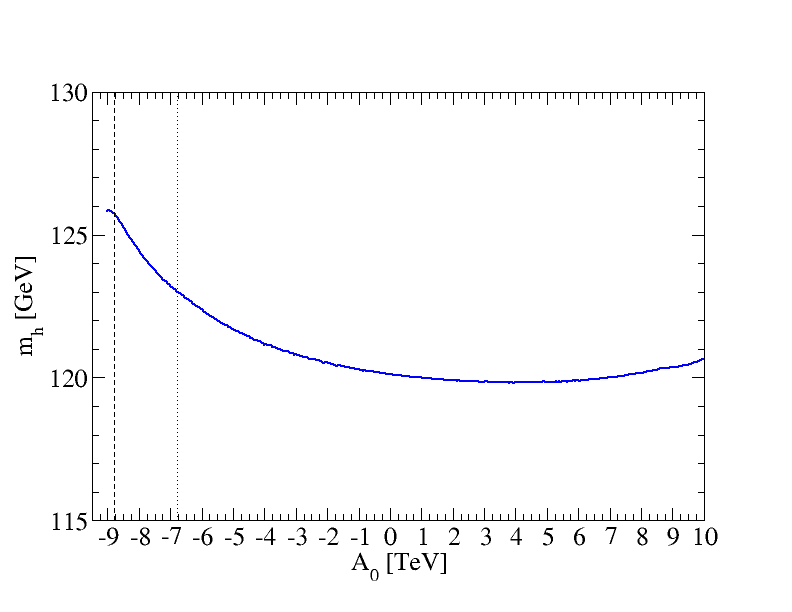}
 \includegraphics[clip, width = 0.49 \textwidth]{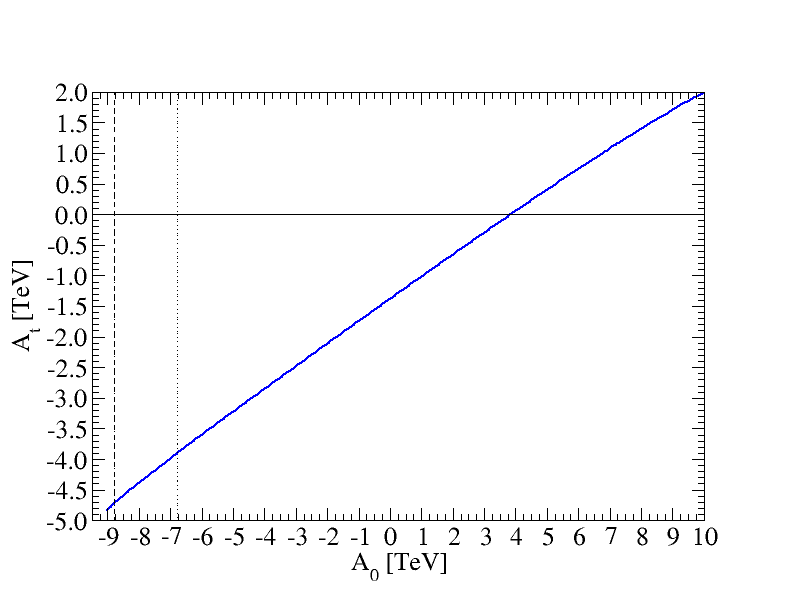}
 \includegraphics[clip, width = 0.49 \textwidth]{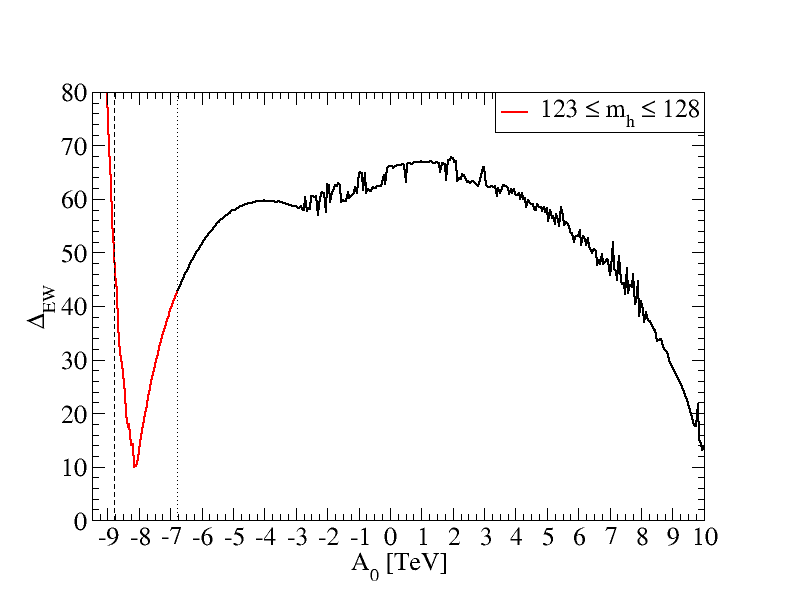}
\caption{In {\it a}), we plot third generation sparticle masses
vs. $A_0$ for an RNS benchmark with $m_0=5$~TeV, $m_{1/2}=900$~GeV,
 $\tan\beta =10$ and with $\mu=125$~GeV and $m_A=1$~TeV. In {\it b}), we
 plot the corresponding value of $m_h$ and in {\it c}) we plot $A_t({\rm
 weak})$ while in {\it d}) we plot $\Delta_{\rm EW}$. The dashed
 vertical line denotes the current lower limit on $m_{\tilde{t}_1} \gtrsim
 850$~GeV from ATLAS top squark searches \cite{atlas_stop_run2} and left of
 the dotted vertical line denotes where $m_h > 123$ GeV. The red-shaded
 part corresponds to $123~{\rm GeV} \leq m_h \leq 128$~GeV. 
}
\label{fig:mvsA0}
\end{center}
\end{figure}

In Fig.~\ref{fig:mvsA0} frame {\it a}), we plot the values of the various
third generation sfermion masses versus variation in the $A_0$ parameter.
It is seen that for $A_0\sim 0$, then the various sfermion masses 
range between 3 and 5 TeV. As $A_0$ becomes large positive or negative, 
the $A_{t,b,\tau}$ contributions to the MSSM RG equations tend to drive
the soft masses $m_{Q_3}^2$ and $m_{U_3}^2$ to lower values due to the 
$X_t$ (Eq.~\eqref{eq:Xt}) contribution to the RG running, 
which is amplified by the large top-quark Yukawa coupling $f_t$. 
The $\ttau_{1,2}$ and $\tb_2$ mass values hardly change since their RG
equations include $X_\tau$ and $X_b$ which are only amplified by the much 
smaller $\tau$- and $b$-Yukawa couplings.
Also, the large $A_t$ term causes large mixing in the top squark sector which
enhances the splitting of the stop eigenstates. Only the value of
$m_{\tst_1}$ is driven to sub-TeV values for $A_0\alt -8.8$ TeV.

In Fig.~\ref{fig:mvsA0}{\it b}), we show the value of $m_h$ vs. $A_0$. 
The Higgs mass at one loop is given by
\be
m_h^2\simeq m_Z^2\cos^2 2\beta +\frac{3g^2}{8\pi^2}\frac{m_t^4}{m_W^2}\left[
\ln\frac{m_{\tst}^2}{m_t^2}+\frac{x_t^2}{m_{\tst}^2}\left(1-\frac{x_t^2}{12m_{\tst}^2}\right)\right],
\ee
where now $x_t=A_t-\mu \cot\beta$ and $m_{\tst}^2\simeq m_{Q_3}m_{U_3}$. 
For a given value of $m_{\tst}^2$, this expression is maximal for 
large mixing in the stop sector with $x_t^{\rm max}=\sqrt{6}m_{\tst}$. 
We see from the plot that $m_h$ is maximal for large negative $A_0$. 
This is because the {\it weak scale} value of $A_t$ is large negative
leading to large mixing in the stop sector. For large positive $A_0$, then
the value of $A_t$ largely cancels against gauge contributions in the 
$A_t$ running so $A_t$ runs to small values at the weak scale 
leading to small mixing and too small a value of $m_h$: see 
Fig.~\ref{fig:mvsA0}{\it c}). 

In frame~\ref{fig:mvsA0}{\it d}), we show the calculated value of 
$\Delta_{\rm EW}$. Here, we see that $\Delta_{\rm EW}\sim 60$ 
for $A_0\sim 0$, but for this value of $A_0$, 
the value of $m_h$ is too small. For $A_0\alt -7$~TeV, then we have
large mixing leading to $m_h\sim 125$~GeV (shown by the red-shaded part
of the curve), but also some suppression in the $\Sigma_u^u(\tst_{1,2}
)$ values leading to very natural solutions with $\Delta_{\rm EW}\sim 10$. 
For $A_0\agt +8$~TeV, then
$\Delta_{\rm EW}$ drops below 30, but unfortunately $m_h$ is too low at
$\sim 120$~GeV.

\section{Naturalness and the $m_{\tst_1}$ vs. $m_{\tz_1}$ plane}
\label{sec:plane}

In this Section, we present a grand overview of the locus of natural
SUSY models in the $m_{\tst_1}$ vs. $m_{\tz_1}$ mass plane. This plane
was initially proposed as a template for top squark searches in
Ref.~\cite{Baer:1994xr} and has now served for several years to give a
panoramic view of top squark search results in various simplified models
from LHC data.

\begin{figure}[tbp]
\begin{center}
 \includegraphics[clip, width = 0.8 \textwidth]{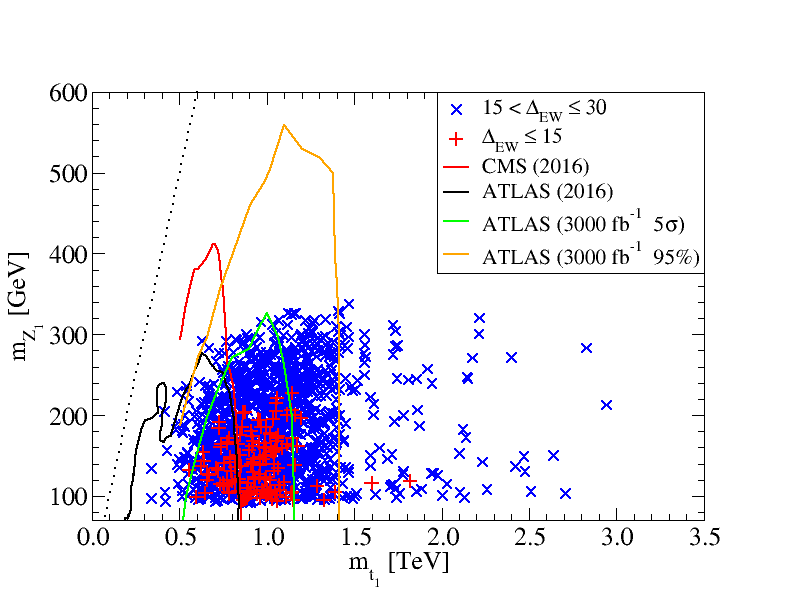}
\caption{The $m_{\tst_1}$ vs. $m_{\tz_1}$ mass plane for 
SUSY with radiatively-driven naturalness and $\Delta_{\rm EW}<15$ (red) 
and 30 (blue). The dotted line denotes the compressed region 
where $m_{\tst_1}=m_{\tz_1}$.
}
\label{fig:mt1mz1}
\end{center}
\end{figure}

In Fig.~\ref{fig:mt1mz1}, we present the results of the scan over NUHM2
parameter space from Ref.~\cite{upper} where upper bounds on
sparticle masses were derived from requiring not-to-large 
values of $\Delta_{\rm EW}$.
The scan values were $m_0:0$--20~TeV, $m_{1/2}:0.3$--3~TeV, $-3<A_0/m_0
<3$, $\mu :0.1$--1.5~TeV, $m_A:0.15$--20~TeV and $\tan\beta :3$--60. It
was required that  
1. electroweak symmetry to be radiatively broken, 
2. the $\tz_1$ was LSP, 
3. the lightest chargino obeyed the LEP2 limit $m_{\tw_1}>103.5$ GeV, 
4. LHC8 bounds on $m_{\tg}$ and $m_{\tq}$ were respected and 
5. $m_h=125\pm 2$ GeV.  
From the Figure, we see that solutions with $\Delta_{\rm EW}<15$ are
clustered with $m_{\tst_1} = 0.6$--1.3~TeV while if we allow for
$\Delta_{\rm EW}<30$ then $m_{\tst_1}$ can range up to 3 TeV.\footnote{
Let us compare this result with that obtained in
Ref.~\cite{Han:2016xet}. The analysis presented in
Ref.~\cite{Han:2016xet} shows that $\delew < 30$ gives an upper bound on
the mass of $\tst_1$ as $m_{\tst_1} \lesssim 1.6$~TeV, which is much
lower than our result. This apparently severe bound results from the
different strategy of the parameter scan, which turns out to be more
restricted than ours. For example, they scan
parameters in the ranges of $100~{\rm GeV} \leq m_{\widetilde{Q}_{3L},
\widetilde{U}_{3R}} \leq 2.5$~TeV and $1~{\rm TeV} \leq A_t \leq
3$~TeV at the weak scale. As can be seen from Fig.~\ref{fig:mvsA0}{\it
c}), however, $A_t < -7$~TeV can give a very small value of $\delew$,
which is out of the range of the parameter scan in
Ref.~\cite{Han:2016xet}. Top-squark masses can also be as large as
$\sim 3$~TeV for $\delew < 30$. 
}
The black-dotted line shows where $m_{\tst_1}\sim m_{\tz_1}$ which is
the compressed region, in which laborious searches for top squark
production are taking place. Notice that essentially no highly natural
solutions lie in this region. It is also important to note that the LSP
is mainly higgsino-like in this region in order to satisfy naturalness
with low $\Delta_{\rm EW}$.

We also present for comparison several
search contours from the ATLAS collaboration. The region within the
solid-black contour represents the area ruled out by current ATLAS
searches at LHC13 for $pp\to\tst_1\tst_1^*$: for $\tst_1\to t\tz_1$ or
$bW\tz_1$ \cite{atlas_stop_run2, ATLAS:2016ljb, ATLAS:2016xcm} and for
$\tst_1\to c\tz_1$ \cite{Aaboud:2016tnv}. These search results range up
to $m_{\tst_1}\sim 850$~GeV which covers only a fraction of the expected
range from natural SUSY. We note, however, that some of these limits might
be significantly relaxed in the present case as they are obtained on the
assumption of a specific decay channel in a simplified setup. For
example, the limit from the ATLAS one lepton, jets plus missing energy
search \cite{ATLAS:2016ljb}, in which all of the produced top squarks are
assumed to decay into $t\tz_1$, would be relaxed since the $\tst_1\to 
t\tz_{1,2}$ decay branch is about 50\% in the natural SUSY parameter
space, as we will see below. Considering this, we also show in
Fig.~\ref{fig:mt1mz1} as the red contour the limits presented in
Ref.~\cite{Han:2016hgr}, which are obtained by recasting the CMS
top-squark mass limits \cite{CMS:2016hxa} for models with light
higgsinos; the resultant upper bound on the top-squark mass is again
found to be about $850$~GeV.

In Fig.~\ref{fig:mt1mz1}, we show as well projected contours of what HL-LHC
can achieve via the top-squark search in the 0-lepton channel; the
$5\sigma$ discovery and 95\% CL exclusion contours with 3000~fb$^{-1}$
integrated luminosity data are shown in the green and orange solid
lines, respectively \cite{ATLAS:2013hta}. Here, we see that HL-LHC with
3000~fb$^{-1}$ of integrated luminosity may be able to probe up to
$m_{\tst_1}\sim 1.4$~TeV. This can be compared with a recent theory
study\cite{Kim:2016rsd} finding HL-LHC may probe top squark pair
signatures to $m_{\tst_1}\sim 1.4$ TeV. A combination of the 0-lepton
and 1-lepton search results may further push its reach by $\sim
50$~GeV \cite{ATLAS:2013hta}. In either case, HL-LHC probes
perhaps less than half the natural SUSY parameter space via top squark
pair searches. 

Before concluding this section, we comment on the excess events
observed in the ATLAS top-squark searches based on the one lepton, jets
plus missing energy final states \cite{ATLAS:2016ljb}, where
$2.2\sigma$, $2.6\sigma$, and $3.3\sigma$ excesses are observed in the
signal categories, {\tt SR1}, {\tt bC2x\_diag}, and {\tt DM\_low},
respectively. As discussed in Ref.~\cite{Han:2016hgr}, these excesses
may be explained with a top squark with a mass of $\lesssim 750$~GeV and
light higgsinos with masses of $\lesssim 200$~GeV. However, such
parameter region has already been excluded by other searches
\cite{atlas_stop_run2, CMS:2016hxa} as shown in Fig.~\ref{fig:mt1mz1},
and thus these excesses are not accounted for in the present
setup.\footnote{We however note that by considering the bino LSP case
with light higgsinos, we may explain the excesses without conflicting
with other limits, as discussed in Ref.~\cite{Han:2016hgr}.  }

\section{The branching fraction ${\rm BF}(b\to s\gamma)$ vs. $m_{\tst_1}$}
\label{sec:bsg}

Here, we examine expectations for the rare branching fraction ${\rm
BF}(b\to s\gamma )$ which takes place via $Wt$ loops in the SM and via
$\tst_i\tw_j$ and $bH^+$ loops in SUSY \cite{Bertolini:1990if,vb_bsg} (other SUSY loops
also contribute but typically with much smaller amplitudes). The SM
value for this decay is found to be \cite{Misiak:2015xwa} ${\rm BF}(b\to
s\gamma )=(3.36\pm 0.23)\times 10^{-4}$ which is to be compared to 
the recent Belle measurement\cite{Belle:2016ufb} that ${\rm BF}(b\to s\gamma )=(3.01\pm 0.22)\times
10^{-4}$. For the SUSY ${\rm BF}(b\to s\gamma )$ calculation, we use the
NLO results from \cite{bsg} which is encoded in Isatools \cite{isajet}. 

\begin{figure}[tbp]
\begin{center}
 \includegraphics[clip, width = 0.8 \textwidth]{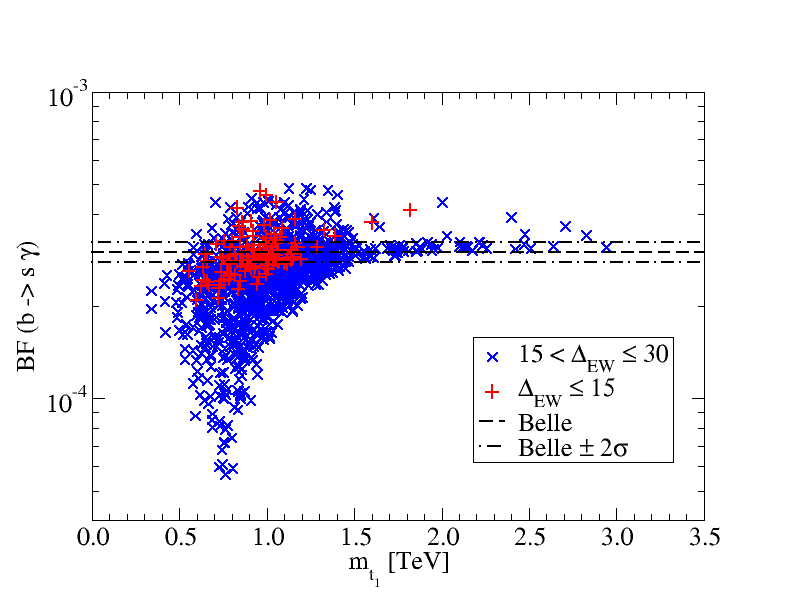}
\caption{Plot of ${\rm BF}(b\to s\gamma )$ vs. $m_{\tst_1}$ for 
SUSY with radiatively-driven naturalness and 
$\Delta_{\rm EW}<15$ (red) and 30 (blue).
}
\label{fig:bsg}
\end{center}
\end{figure}

In Fig.~\ref{fig:bsg}, we show the predicted value of ${\rm BF}(b\to
s\gamma )$ from our scan over NUHM2 model parameters for points
satisfying $\Delta_{\rm EW}<15$ (red) and 30 (blue) versus $m_{\tst_1}$. 
The various constraints from above, including LHC search and
compatibility with $m_h$, are included. We also indicate the Belle
central value and $\pm 2\sigma$ bounds by the dashed and dot-dashed
lines, respectively. From the plot, we see a large deviation between the
predicted and measured values of ${\rm BF}(b\to s\gamma )$ for light
$m_{\tst_1}$ values. Especially noteworthy is that {\it no} values of
${\rm BF}(b\to s\gamma )$ lie within the $\pm 2\sigma$ measured band for
$m_{\tst_1}<500$ GeV. Recall that this range of stop masses 
is often considered generally natural \cite{oldnsusy} before amending the
calculations of $\Delta_{\rm HS}$ and $\Delta_{\rm BG}$. As $m_{\tst_1}$
increases, the predicted range of ${\rm BF}(b\to s\gamma )$ rises
asymptotically to be within the measured range: this occurs especially
for $m_{\tst_1}>1.5$ TeV. The intermediate region with $0.5\ {\rm
TeV}<m_{\tst_1}<1.5$ TeV contains points in agreement with the measured
value, where the various $\tst_{1,2}\tw_{1,2}$ amplitudes, which can
occur with either positive or negative values, cancel one-with-another. 
But even in this region of $m_{\tst_1}$ values, the bulk of points tend
to deviate severely from the measured value. This is because the loop
contributions can always be large since the higgsino-like charginos and
stops are both light. From examining the confrontation between predicted
and measured values of ${\rm BF}(b\to s\gamma )$, it comes as no
surprise that light stops have yet to be detected at LHC. 
%

\section{Top squark production and decay at LHC}
\label{sec:prod}

%
\begin{figure}[tbp]
\begin{center}
 \includegraphics[clip, width = 0.8 \textwidth]{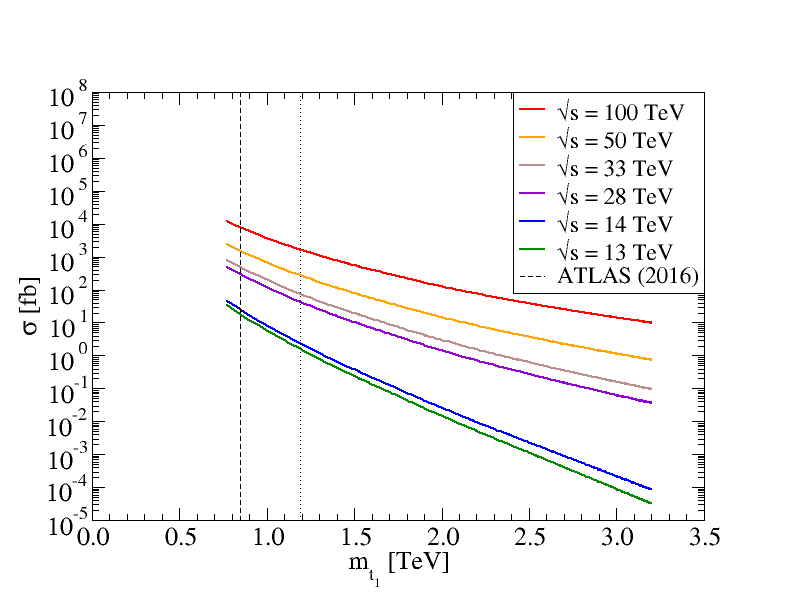}
\caption{NLO top squark pair production cross section vs. 
$m_{\tst_1}$ for $\sqrt{s}=13$, 14, 28, 33, 50 and 100~TeV. 
The dashed vertical line denotes the current lower limit on $m_{\tilde{t}_1} \gtrsim 850$~GeV 
from ATLAS top squark searches and the dotted vertical line denotes 
the projected reach of HL-LHC.
}
\label{fig:sigt1}
\end{center}
\end{figure}

In this section, we consider top squark pair production and decay rates 
at the LHC. Top squark pair production proceeds dominantly through
the QCD $gg$ and $q\bar{q}$ annihilation channels. 
The NLO production rates for LHC with $\sqrt{s}=13$ and 14~TeV are calculated
using Prospino \cite{prospino} and shown in Fig.~\ref{fig:sigt1} versus
top squark mass $m_{\tst_1}$. 
We also show production rates for future proposed $pp$ colliders
operating with $\sqrt{s}=28$, 33, 50 and 100 TeV.
The vertical dashed line shows the
approximate locus of the ATLAS/CMS bounds on $m_{\tst_1}$ from searches
within the context of simplified models with a low value of
$m_{\tz_1}$. 
The dotted vertical line denotes the projected reach of HL-LHC for top-squarks. 
We see that the total production cross section for
$m_{\tst_1}\sim 850$~GeV at LHC14 are in the $10$--20~fb range. By moving up to
$m_{\tst_1}\sim 1200$~GeV, the cross section drops by about an order of
magnitude to about $1$~fb. At $m_{\tst_1}\sim 1.6$~TeV, $\sigma (pp\to
\tst_1\tst_1^*)$ drops by another order of magnitude to about
0.1~fb. These total cross sections may be compared to the upper limit on
$m_{\tst_1}$ from requiring $\Delta_{\rm EW}<30$ whereupon
$m_{\tst_1}<3$~TeV is required. For such large values of $m_{\tst_1}$,
the total cross sections are in the $10^{-3}$~fb range. Probing such
massive top squarks will likely require an LHC energy upgrade 
(HE-LHC with $\sqrt{s}\sim 28$--33~TeV) or else a future circular
collider (FCC) with $\sqrt{s}\sim 50$--100~TeV
\cite{Cohen:2014hxa,pp100, Golling:2016gvc}.

\begin{figure}[tbp]
\begin{center}
 \includegraphics[clip, width = 0.7 \textwidth]{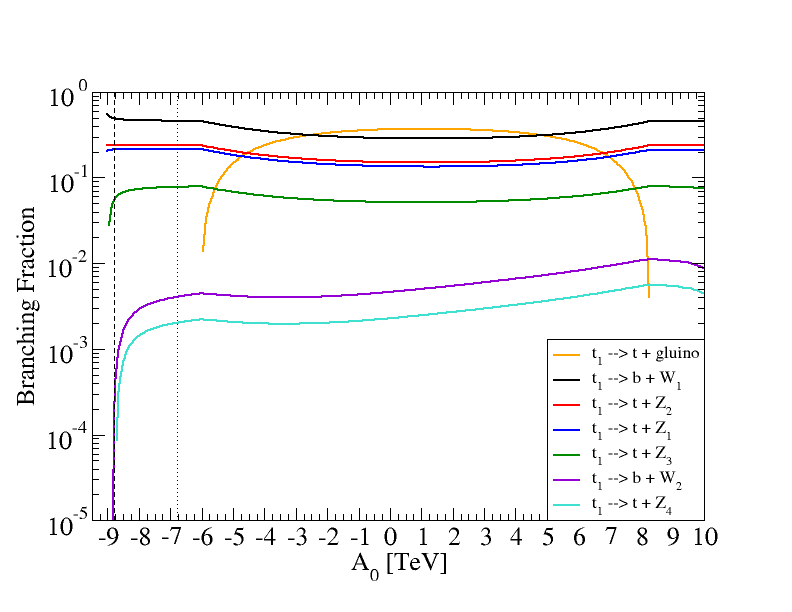}
 \includegraphics[clip, width = 0.7 \textwidth]{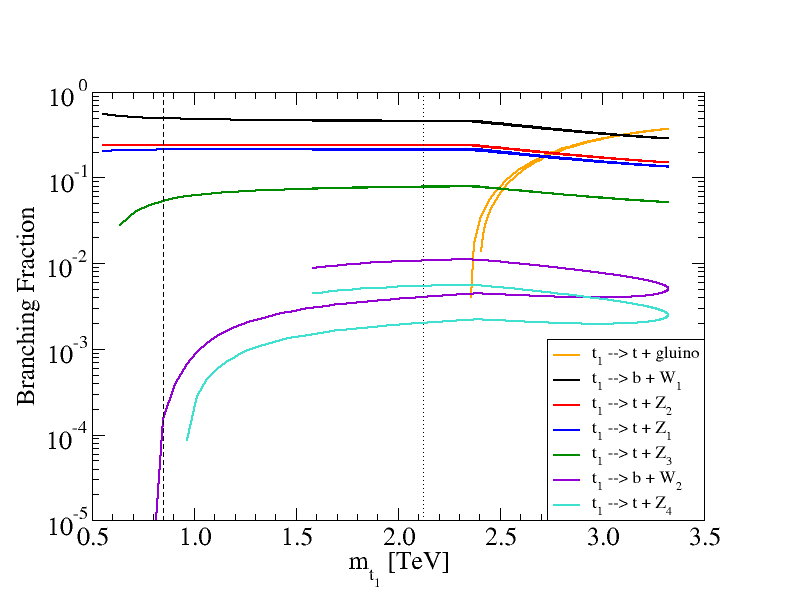}
\caption{Top squark branching fractions vs. {\it a}) $A_0$ 
and {\it b}) $m_{\tst_1}$ along the RNS model-line. 
Left of the dotted vertical line is where $m_h>123$ GeV 
while left of the dashed vertical denotes where $m_{\tst_1}<850$ GeV.
}
\label{fig:BFt1}
\end{center}
\end{figure}

In Fig.~\ref{fig:BFt1}{\it a}), we show the expected top squark 
branching fractions versus $A_0$ along the top squark model line. 
The branching fractions are from Isajet~\cite{isajet}.
In the plot, the black curve denotes ${\rm BF}(\tst_1\to b\tw_1 )$
where for our model line $\tw_1$ is the lighter, mainly higgsino-like, 
chargino. This mode occurs at the $\sim 50\%$ rate and is rather model
independent (within the context of natural SUSY with light higgsinos).
The $\tw_1$ further decays via 3-body mode into $\tw_1\to
f\bar{f}'\tz_1$ where $\tz_1$ is the higgsino-like LSP. Since
$m_{\tw_1}-m_{\tz_1}$ (and $m_{\tz_2}-m_{\tz_1}$) are $\sim 10$--20~GeV,
most of the decay energy goes into making the $\tz_1$ rest mass and is
undetected. The $f\bar{f}'$ energy is rather soft leading to a few soft tracks. 
Thus, both the $\tw_1$ and $\tz_2$ are only quasi-visible. 
Meanwhile, the $b$-jet from $\tst_1\to b\tw_1$ decay may be quite hard, 
typically in the hundreds of GeV.

The red and blue curves denote the ${\rm BF}(\tst_1\to t\tz_2 )$ and
${\rm BF}(\tst_1\to t\tz_1)$ respectively. Both these branching fractions come in at the 20--25\% level thus covering the bulk of the remaining decays. 
While the $\tz_1$ is invisible (it presumably comprises a portion of the
dark matter), again the $\tz_2$ and $\tw_1$ are quasi-visible. Meanwhile,
the top quarks are produced at large $p_T$ and also their rest mass
leads to energetic decay products. In addition, there is a non-negligible 
decay rate $\tst_1\to t\tz_3$ where $\tz_3$ is bino-like and yields
visible decays. These decays occur at the few percent
level. Furthermore, $\tst_1$ decays into wino-like $\tw_2$ and $\tz_4$
can occur but at the sub-percent level. 
The dip in branching fractions at the center of the plot is due to 
turn on of $\tst_1\to t\tg$.
%

In Fig.~\ref{fig:BFt1}{\it b}), we show the same branching fractions versus
$m_{\tst_1}$ along the model line. 
The branching fractions are again seen to be rather model independent 
except for $m_{\tst_1}\sim \mu$ (in the excluded range)
where the decays into top-quarks become kinematically forbidden. 
The branching fractions in this plot are double-valued since certain
top squark mass values can occur for both large positive and large negative
values of $A_0$. 
These mainly affect the tiny branching fractions into wino-like electroweakinos.
%

\section{Prospects for top squark discovery at LHC and beyond}
\label{sec:prospects}

The most direct implication of naturalness is the existence of
light higgsinos of mass $m_{\tw_1,\tz_{1,2}}\sim 100$--300~GeV, 
the lighter the better. Given these expectations on $m({\rm LSP})$, 
the LHC lower bound
$m_{\tst_1}\agt 850$~GeV applies and we expect top squarks to lie in the mass 
range $m_{\tst_1}\sim 850$--3000~GeV at little cost to naturalness. 
This mass range is consistent with expectations from comparing
the predicted ${\rm BF}(b\to s\gamma )$ to its measured value. 
Then, the highly scrutinized $\tst_1 -\tz_1$ degeneracy rarely if ever applies
and we expect instead a rather large $m_{\tst_1}-m_{\tz_1}$ mass difference.
In this case, the top squark branching fraction predictions from
Sec.~\ref{sec:prod} are rather robust: they result over a huge range of
NUHM2 parameter space and also under the natural general mirage
mediation parameter space found in Ref.~\cite{ngmm}.\footnote{The
non-universal gaugino mass models \cite{abe} also predict similar top
squark branching ratios \cite{Abe:2015xva, Kawamura:2016drh}. } We would then expect, quite generally, the following collider signatures to obtain:
\bi
\item A. $\tst_1\tst_1^*\to b\bar{b}+\eslt \ \ \ \sim 25\%$ ,
\item B. $\tst_1\tst_1^*\to b\bar{t},\ \bar{b}t +\eslt \ \ \ \sim 50\%$ ,
\item C. $\tst_1\tst_1^*\to t\bar{t}+\eslt \ \ \ \sim 25\%$ .
\ei
These signatures should be accompanied by the usual initial state 
radiation plus perhaps additional semi-soft tracks from associated 
light higgsino $\tw_1$ and $\tz_2$ decays.

The first signal channel A. includes rather hard $b$-jets plus hard $\eslt$
and should be plagued by backgrounds including $b\bar{b}Z$ 
production where $Z\to \nu\bar{\nu}$. One might 
create distributions using the $m_{T2}$ variable applied to the $b\bar{b}+\eslt$
final state to try to extract a kinematic upper edge which could yield 
an estimate of the top squark mass.

For signal channel B., we expect a hard $t$-jet along with a hard $b$-jet
and $\eslt$. This channel would include $b\bar{b}+\eslt$ 
along with an added $W\to f\bar{f}'$ where in the case of hadronic $W$ decays, 
the $W$ mass may be reconstructed. The dominant backgrounds
would include $t\bar{t}$ production, $Wb\bar{b}$ production and 
$WZ$ production where $Z\to\nu\bar{\nu}$ and $g\to b\bar{b}$, 
single top production and $tbZ$ production.
This ``mixed top-squark decay channel'' has previously been emphasized
by Graesser and Shelton\cite{Graesser:2012qy}.

Signal channel C. contains a hard $t\bar{t}$ pair plus large $\eslt$. 
Major backgrounds would include $Zt\bar{t}$ production. The hard $t$-jets
may benefit from a top-tagger\cite{Kaplan:2008ie}.

A credible semi-simplified model could be presented in the $m_{\tst_1}$ 
vs. $m(\text{higgsino})$ mass plane where the several dominant decay branching 
fractions would be allowed to take place.
Physically, this is what is expected to happen and one would then include
the dominant mixed decay mode where one $\tst_1$ decays to $b\tw_1$ 
while the other decays to $t\tz_{1,2}$.

Finally, we comment on indirect searches for top squarks at the
LHC. Since $m_{\tst_1}:0.85$--3~TeV is predicted in the radiatively-driven
natural SUSY, one may expect that its signature can be probed
indirectly via the precise measurements of the Higgs decay branching
ratios, as top squarks affect the $h\to \gamma \gamma$ and $h\to gg$
decay channels at one-loop level. As it turns out, however, the deviations of
these decay branches from the SM prediction are too small to be detected
even at the HL-LHC \cite{Bae:2015nva}. This observation again leads to
the conclusion that future colliders such as ILC or an energy upgraded LHC
are required for a thorough coverage of (just the top-squark sector of) 
natural SUSY.

\section{Conclusions}
\label{sec:conclude}

In this paper we have re-examined the phenomenology of top squarks
expected from natural SUSY. We first noted that older expectations of 
very light top squarks based on requiring small 
$\delta m_{H_u}^2/m_h^2$ are technically flawed in that they neglect the contribution
of $m_{H_u}^2$ to its own running. By properly including this contribution, 
then the $\Delta_{\rm HS}$ measure reduces to $\Delta_{\rm EW}$. 
The $\Delta_{\rm EW}<30$ requires light higgsinos $\sim 100$--300~GeV while 
much heavier top squarks $m_{\tst_1}\sim 0.85$--3~TeV 
are allowed at little cost to naturalness. 
In the latter case, the radiative corrections to $m_{H_u}^2$ aid in driving 
it from large unnatural high scale values to natural values at the weak scale---a situation known as radiatively-driven natural SUSY or RNS.
For the case of BG naturalness, if $\Delta_{\rm BG}$ is evaluated in multi-soft-parameter effective theories, 
then one obtains an overestimate of fine-tuning as compared to the calculation 
for a more fundamental theory wherein the soft terms are all correlated. 
In the latter case, $\Delta_{\rm BG}$ reduces to $\Delta_{\rm EW}$. 

Using $\Delta_{\rm EW}$, it is found that current LHC top squark 
search constraints have probed only a fraction of the allowed
$m_{\tst_1}$ vs. $m_{\tz_1}$ parameter plane. The compressed region, which has been heavily searched, admits few or no solutions. Further, values of 
$m_{\tst_1}<500$~GeV lead to typically large deviations in ${\rm BF}(b\to s\gamma )$.

%
\begin{figure}[tbp]
\begin{center}
 \includegraphics[clip, width = 0.7 \textwidth]{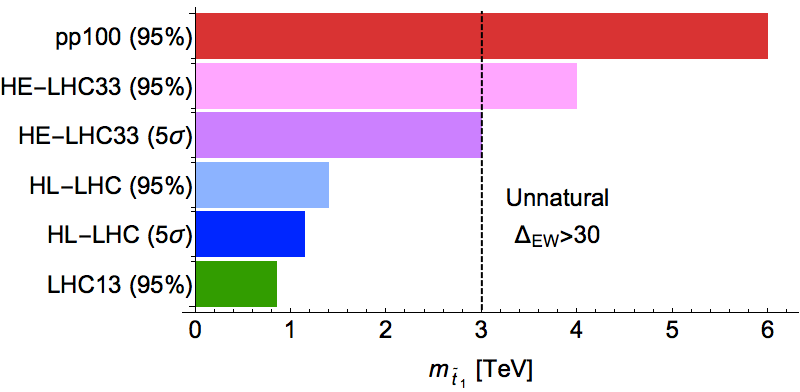}
\caption{Current limits on top-squarks along with projected 
discovery and exclusion reaches of future possible colliders.
}
\label{fig:bar}
\end{center}
\end{figure}

Top squark production and decay rates are calculated in natural SUSY and lead to comparable  mixtures 
of $\eslt$ plus $b\bar{b}$, $t\bar{t}$ and $tb$ signatures.
It is emphasized that a semi-simplified model containing the major 
admissible final states would be most helpful to truly constrain the natural SUSY 
parameter space or to discover top squarks. 
Nonetheless, plenty of 
perfectly natural SUSY solutions exist with $m_{\tst_1}$ values  well beyond the
reach of HL-LHC. To probe the entire expected natural SUSY top squark 
parameter space will likely require an energy upgrade of LHC to the
$\sqrt{s}\sim 28$--33~TeV regime.
To this end, in Fig.~\ref{fig:bar}, we show the current exclusion limit 
on top squark masses from ATLAS/CMS for a light $\tz_1\sim 100$--200~GeV. 
We also show the HL-LHC projected reach and exclusion limits for 3000 fb$^{-1}$
of integrated luminosity \cite{ATLAS:2013hta} along with the projected reach of 
future $pp$ colliders with $\sqrt{s}=33$ and 100 TeV\cite{Gershtein:2013iqa}. 
In contrast to common notions, the display shows that HL-LHC has a 
very limited reach for natural SUSY in the top-squark pair production channel.
Even if no top-squark signal is seen at HL-LHC, then there will be little impact on 
excluding natural SUSY (other channels such as same-sign diboson or soft dilepton plus jets appear
more lucrative to HL-LHC)\cite{Baer:2016usl}. 
However, an energy upgrade to HE-LHC with $\sqrt{s}=33$ TeV
will have a $5\sigma$ discovery reach to $m_{\tst_1}\sim 3$ TeV and a 95\% CL exclusion reach
to 4 TeV. 
Such a reach will either discover or exclude natural SUSY in the top squark
sector. 
We also show the Snowmass projected reach\cite{Gershtein:2013iqa} 
for top-squark pairs for a 100 TeV collider.
Such a machine is projected to probe up to $m_{\tst_1}\sim 6$ TeV. 
This reach probes beyond a 33 TeV machine only further 
into unnatural regions of parameter space.

\section*{Acknowledgments}

We thank X. Tata for comments on the manuscript.
This work was supported in part by the US Department of Energy, Office of High
Energy Physics.
%

%
\end{document}